\documentclass[letter,10pt]{sig-alternate-10pt}

\usepackage{times,graphicx,subfigure, amsmath}
\usepackage{epstopdf,comment,color,url}
\usepackage{algorithmic}
\usepackage{algorithm}
\usepackage{multirow}
\definecolor{Red}{RGB}{255,0,0} 
\definecolor{Blue}{RGB}{0,0,255} 
\definecolor{Green}{RGB}{0,255,0} 

\begin{document}
\sloppy
\title{Characterizing Docker Overhead \\ in Mobile Edge Computing Scenarios}

\numberofauthors{5} 
\author{
Giuseppe Avino, Marco Malinverno, Francesco Malandrino,\\ Claudio Casetti,
Carla-Fabiana Chiasserini\\
Politecnico di Torino, Italy\\
C.so Duca degli Abruzzi 24, 10129 Torino, Italy
}
\maketitle


%
\begin{abstract}
Mobile Edge Computing (MEC) is an emerging network paradigm that provides cloud and IT services at the point of access of the network. Such proximity to the end user translates into ultra-low latency and high bandwidth, while, at the same time, it alleviates traffic congestion in the network core. Due to the need to run servers on edge nodes (e.g., an LTE-A macro eNodeB), a key element of MEC architectures is to ensure server portability and low overhead. A possible tool that can be used for this purpose is Docker, a framework that allows easy, fast deployment of Linux containers. This paper addresses the suitability of Docker in MEC scenarios by quantifying the CPU consumed by Docker when running two different containerized services: multiplayer gaming  and video streaming. Our tests, run with varying numbers of clients and servers, yield different results for the two case studies: for the gaming service, the overhead logged by Docker increases only with the number of servers; conversely, for the video streaming case, the overhead is not affected by the number of either clients or servers.

\end{abstract}
\section{Introduction}
It is expected that 5G networks will be asked to handle a wealth of applications with a large set of requirements. The main actors of the ``5G revolution'', along with mobile network operators, are \textit{vertical industries}  such as entertainment/media companies, car manufacturers, eHealth operators, etc. The applications and services that verticals are likely to deploy will have  high  availability and reliability requirements, and they will need to be provided with low power consumption and, more often than not, with stringent latency. The design and implementation of the 5G platform that must cater for such demanding services, and on top of which  network and IT-related functionalities must reside, will need to be highly innovative. At a minimum, it will have to support 
novel functional split and placement methods, so as to enable the joint optimization of IT and networking resources and optimally serve vertical industries requirements. 

In this scenario, the Mobile Edge Computing (MEC) paradigm is set to play a crucial role.  Deploying services in a decentralized fashion at the edge allows more closely-knit interaction between service provision and end users, drastically reducing latency and  boosting service availability.
Decentralization of service distribution, however, comes with a price: {\em duplication of resources and overhead increase}, which should be accurately quantified  based  on the platform of choice.
The use of virtual machines and lightweight containers in the provision of services is the approach which the IT industry is currently investing on. The most widely used container framework is arguably Docker,  promising  to guarantee high portability, scalability and streamlined collaboration between developers and 
operators~\cite{docker:web}.

Among the multitude of services that can be exposed in MEC scenarios, we selected video streaming and multiplayer gaming, due to their steadily increasing growth and impact. It is a well-established fact that video streaming is one of  the biggest contributors to Internet traffic: it involves an amount of data greater than a picture or a web page, and countless videos are daily streamed on the Internet (YouTube videos, embedded Facebook clips or Netflix series, in addition to videoconferencing tools such as Skype, Join.me or Google Hangouts). According to the latest Cisco statistics, videos represent 70\% of all the Internet traffic and in 2020 such a share could hit 90\% \cite{cisco}.
The online gaming industry is continuously on the rise too: in 2016 it reached the staggering value of \$100 billions in revenues, of which more than  37\%  is represented either by tablets or smartphones.
Furthermore, projections see the game industry passing the \$120 billions mark by 2020, making it   very likely that significant money and effort will be invested in the development of new applications and technologies  for mobile gaming\cite{game-article}. 

In addition to their popularity and widespread adoption, we selected video streaming and gaming as they are representative of service models with significantly different requirements. Video streaming typically requires low computational effort at the server side but a fairly large amount of transmitted data, mainly traveling in one direction. Gaming, instead, implies a much higher utilization of CPU resources and a smaller number of packet transmissions (which depends on the level of interaction between the players in the gaming session), travelling in both directions. 

Taking the above services as case studies, the goal of this work is to measure and characterize the Docker overhead implied by their implementation within containers deployed at the edge of the mobile network infrastructure. 
Specifically, by deploying popular services on real hardware, we  study the CPU consumption logged by Docker to run the containers, and discuss the different factors influencing it, e.g., the number of servers and clients. Furthermore, we compare the Docker CPU overhead with the CPU consumption of dockerized applications, in order to gain insight on the price to pay to containerize video streaming and gaming  services.

\section{Related works}
Containerization systems for service delivery in 5G networks is a widely studied topic. 
In particular, works such as \cite{dockevaluation} evaluate Docker as a platform for Edge Computing concluding that  it provides fast deployment, small footprint and good performance.
Other studies, e.g., \cite{AaaS}, claim that distributed service delivery will be one of the hottest topics in 5G networks, and demonstrate the efficiency of Docker as containerization system in the implementation of Virtual Network Functions. Furthermore, Linux containers have been widely analyzed and compared against Virtual Machines under multiple aspects: \cite{comparison} explores the performance of traditional VM deployments, and compares this approach to the use of Linux containers. In line with our results, \cite{comparison} also demonstrates that containers lead to equal or better performance than VMs in all cases. 

Other works, such as \cite{game}, focus on the advantages of Docker containers rather than VMs, in the deployment of cloud gaming, i.e., architectures where the game is rendered on a distant cloud server and the resulting video stream is sent back to the user. We underline that in our work,  instead, a dockerized multiplayer client-server architecture is analyzed.
The authors of \cite{overlay} investigate Docker in overlay-network mode and in host-network mode for the provisioning of a large-scale streaming system, finding that the advantages in terms of latency in host mode are balanced by the higher stability of an overlay network. 
Although related to their study, our work focuses on characterizing the overhead cost of the containerized solution for different applications.

\section{System scenario and approach}
We consider a MEC scenario where services to mobile users are provided by servers deployed at the edge of the network infrastructure, i.e., close to Wi-Fi access points (AP) or cellular base stations (eNodeB). 
As representative of services exhibiting different characteristics and requirements, we select video streaming and online gaming -- the former being an already very popular application and the latter steadily gaining an increasing market share. 

We then investigate the overhead cost of providing such services through Docker containers. To this end we take an experimental approach and deploy the testbed shown in Figure~\ref{img:testbed}, involving  a node of the network edge and a varying number of servers implemented therein. Then, through extensive measurements, we analyze the overhead exhibited by Docker, and its behavior as the application bit rate and the number of clients vary. 

\subsection{Testbed description}
\begin{figure}[b]
\centering
\includegraphics[scale=0.23]{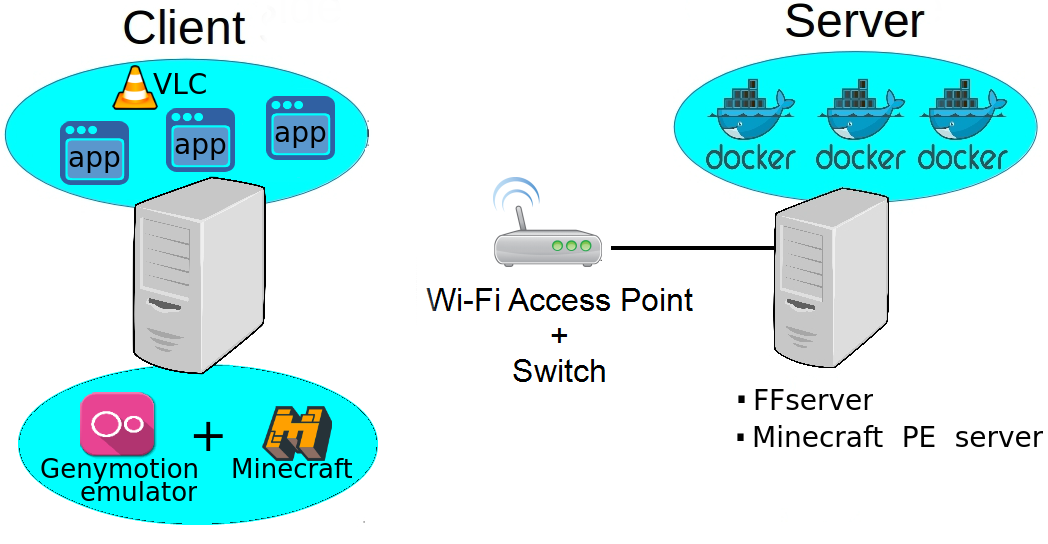}
\caption{Layout of our testbed: clients (left) access the containerized servers through an AP.}\label{img:testbed}
\end{figure}
The video server in our tests is \textit{FFserver} (version 3.1.3) \cite{ffmpeg}, part of the \textit{FFmpeg} framework. FFserver is a multimedia streaming server for both live and non-live content; furthermore,  it supports clients running on multiple platforms, hence the content it provides can be reproduced by any type of device, be it a desktop, a laptop, or a mobile device. The software we use at the client side is \textit{VLC} \cite{vlc}.
As depicted in Figure \ref{img:testbed}, the server instances run through a Wi-Fi AP, which streams video traffic to  VLC instances running on a Linux machine, associated with the AP \footnote{Notice that we could replace Wi-Fi with any other wireless access technology, e.g., LTE, without altering the architecture of our testbed.}.

As far as the dockerized gaming server is concerned, we use \textit{Minecraft Pocket Edition} (version 0.10.5) \cite{minecraft}.
In order to test the behavior of mobile clients, we resort to Android emulators that mimic Android mobile clients. 
The chosen emulator  is \textit{Genymotion}, a powerful virtualization platform capable of emulating many Android devices \cite{genymotion}. In order to be as rigorous as possible, the Android application \textit{FRep} 
is installed in each emulator \cite{frep}. This tool records a sequence of taps on the screen and allows to replicate exactly the same movement pattern over each emulator, thus over each Minecraft client. This is necessary because, as shown by our measurements, mobility patterns of different players correspond to different  CPU loads. Also,  since Android emulators are computationally heavier than simple VLC instances, in the gaming use case up to two different Linux machines are used at the client side.

In both video streaming and gaming tests,  we use different combinations of the number of clients and the number of servers, each of them ranging between 1 and 8 (e.g., 1 server - 4 clients, 4 servers - 8 clients, etc.).  

\begin{figure}[b!]
\centering
\subfigure[Diagram of the Docker Processes Tree]
  {\includegraphics[scale=0.23]{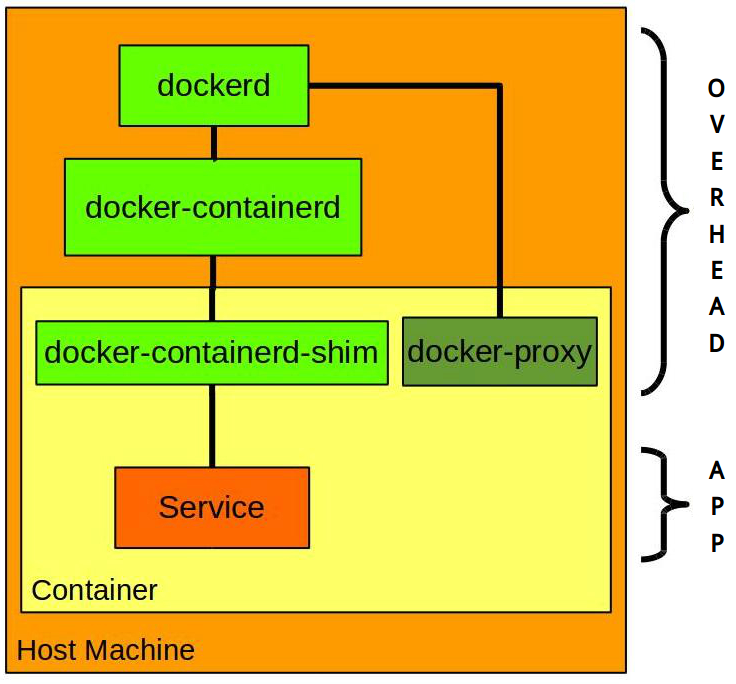}}
\hspace{5mm}
\subfigure[Screenshot of the output of \texttt{pstree}]
  {\includegraphics[width=8.5cm]{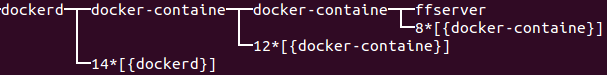}}
\caption{Docker Process Tree} \label{img:tree}
\end{figure}

\section{The Docker overhead}
We first list the processes composing the Docker overhead in subsection ~\ref{sub:processes}, then, in subsection \ref{sub:overhead}, we describe the methodology  used to monitor these processes.

\subsection{The Docker processes tree}
\label{sub:processes}
The Docker overhead is due to several processes whose number varies depending on different factors, e.g., number of containers running in background or foreground, number of exposed ports. Figure \ref{img:tree} shows the Docker processes tree. The top two processes are  persistent, i.e., they are always active no matter whether there are running containers or not; the others are non-persistent, being created as soon as a Docker container is started. 
The persistent processes are the following: 
\begin{itemize}
\item {\tt dockerd}: it is the Docker daemon process. It manages all images, containers and the entire Docker framework. As soon as the daemon is started, it forks the second persistent process, i.e., {\tt docker-containerd}.
\item {\tt docker-containerd}: it manages the lifecycle of containers on the Docker host, \textit{independently} of the Docker daemon, using a shim process (see below). It has been introduced to perform specific maintenance actions on the Docker daemon, such as rebooting or upgrading, without disrupting the normal container execution. 
\end{itemize}

The non-persistent processes are:
\begin{itemize}
\item {\tt docker-containerd-shim}: it is created by {\tt docker-containerd} and it acts as the parent of the container process to facilitate the following operations. First, it allows runtime, low-level components, such as \textit{runC}, to exit after the container is started. In this way, we do not have long-running runtime processes managing the container but we can isolate the dockerized application and the shim process. Secondly, it keeps the STDIO and other file descriptors open for the container in the case where docker-containerd and/or dockerd both terminate. This means that if the shim process is not running, then the parent side of the pipes is closed and the container will exit. Finally,  it allows the container exit status to be reported back to a higher level tool (i.e., to the Docker daemon).  
\item {\tt docker-proxy}: for each port exposed by a container, a docker-proxy process is created. Docker-proxy operates in user space and receives any packet arriving at the specified host port (that the kernel has not dropped or forwarded) and redirects the packet to the container port. Actually, this process is now unused, as it was introduced to get around a limitation exhibited by old kernels, and it has been kept  only for backwards compatibility purposes. Since our Docker host is equipped with an updated kernel, we switched off this process. 
\item {\tt docker}: it is present only if the container runs in foreground as it manages the container user interface. Since in  our case the containers run in background, we do not observe this process in our tests.
\end{itemize}

\subsection{Measuring the Docker overhead}
\label{sub:overhead}
To study the CPU consumed by the processes described above, we parse the files \texttt{/proc/PID/stat}. By retrieving the Process IDentifier in which we are interested, it is possible to gather a great deal of  information. In particular, we are interested in two fields: the \texttt{utime} (user time), i.e., the time length for which a process has been scheduled in user mode, and the \texttt{stime} (system time), i.e., the time the process spends in system mode. Both these fields report measurements expressed in CPU ticks. Generally, in Linux, 1 CPU tick means that the process  occupied the CPU core for 10 ms \cite{cpu}. In order to record the whole CPU usage evolution over time, we sample the \texttt{stat} file once per second. 

Another important element is the amount of data processed by each dockerized application.  The quantity of data exchanged by the network card(s) of the host is stored in the file \texttt{/proc/net/dev}: it contains, for each card, several pieces of information, such as the number of packets sent and received as well as the quantity of bytes transmitted and received. Furthermore, since for every container Docker creates a virtual ethernet card (\textit{veth}), it is possible to monitor the data transferred by each container. 

In summary, sampling each second the   \texttt{stat} and \texttt{dev} files,  for every test it is possible to monitor the  CPU consumption of each process as well as the data processed by each container, second by second.

\section{Experimental results and discussion}

In our testbed, dockerized servers are hosted on a machine with  32-GB RAM memory, an octa-core \textit{Intel Core i7-4790 @ 3.60GHz} processor, and running \textit{Ubuntu 14.04}. Each test lasts for 300 seconds.  
In the video streaming case study, each server streams the same video, with the following features:
\begin{itemize}
\item Format: mpeg;
\item Average Bitrate: 4215 kbps;
\item Resolution: 1280x720 (720p).
\end{itemize}
The images developed to build the two containers are available in Docker Hub at the following repos: \url{https://hub.docker.com/r/giuseppeav/ffmpeg/}, \url{https://hub.docker.com/r/marcomali/minecraft/}.

In the case of gaming, Android emulators are deployed in the laptops used for the tests, connected under the same subnet.

Below, we first fix the number of clients to 8 and vary the number of servers (subsection \ref{fixclient}). Then, we consider a single server with varying number of clients (subsection \ref{fixserver}). Finally, we analyze the impact of  data processed by  containers on the Docker CPU consumption (subsection \ref{data-an}).

\begin{figure}[t!]
\centering
\includegraphics[scale=0.35]{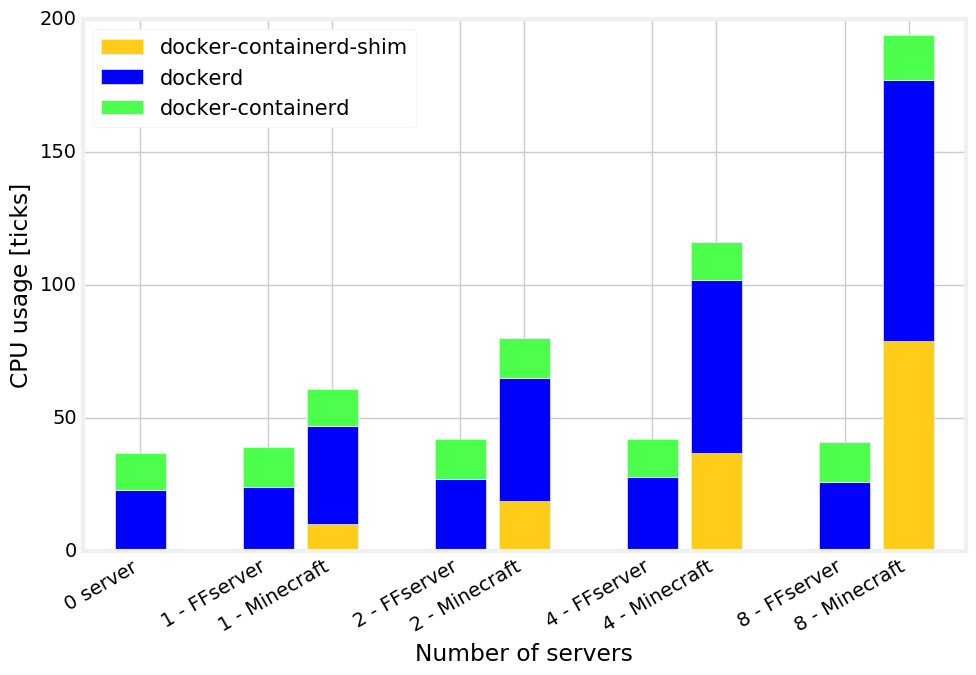}
\caption{Docker overhead for 8 clients and a varying number of servers}\label{img:fixclient1}
\end{figure}  

\subsection{Fixed number of clients} \label{fixclient}
\begin{figure}[b!]
\centering
\subfigure[FFserver]
  {\includegraphics[width=9cm]{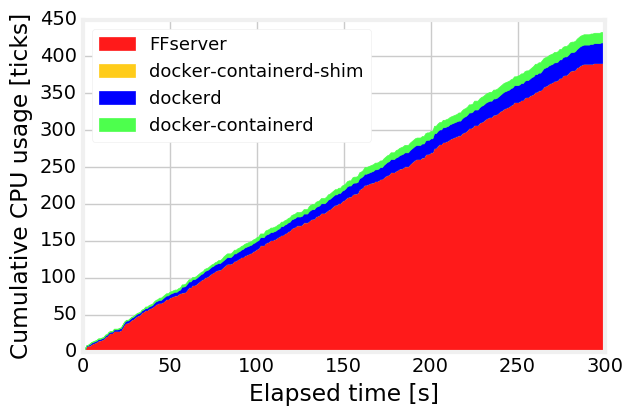}}
\hspace{5mm}
\subfigure[Minecraft]
  {\includegraphics[width=9cm]{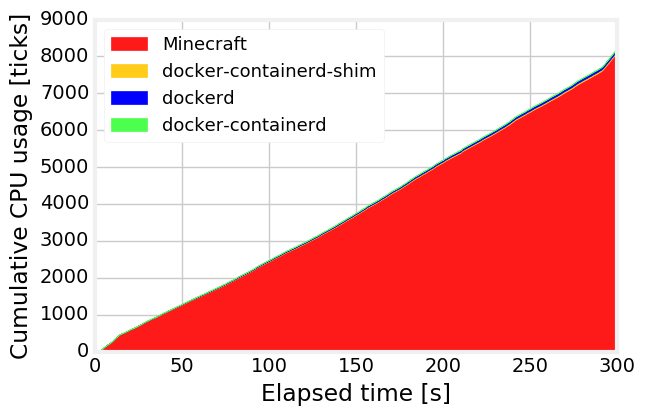}}
\caption{Temporal evolution of the CPU consumption due to the application and to the Docker overhead, in the case of 8 clients and 4 servers} \label{image:fixclient2-3}
\end{figure}
For a fixed number of clients (namely, 8),  intuitively, one could expect that the Docker overhead depends almost linearly with the number of servers.
Figure \ref{img:fixclient1} presents the Docker overhead CPU consumption for 8 clients computed at the end of the test for both video streaming and gaming, along with the results for the case with no running containers, i.e., when only the daemon is active. 

As the number of servers changes, two facts strike our attention. First, for FFserver, the CPU utilization by {\tt dockerd} and {\tt docker-containerd} is not affected by the number of containers (it remains equal to about 40 ticks), and it is essentially the same as the one observed in the absence of any server; furthermore,  the shim processes do not consume any CPU. This implies that the only price to pay to run dockerized FFserver instances is due to the creation and termination  of the containers. Second, for Minecraft, {\tt dockerd}, {\tt docker-containerd} and shim processes all consume CPU, and the amount of CPU used by both shim and {\tt dockerd} does depend on the number of running servers, i.e., the number of containers.

The difference in CPU consumption by the shim process for the two applications can be related to different behaviors of the applications themselves. A prime example is represented by logging operations: the Minecraft game server produces substantially more logs than the video server, and this puts a higher strain on the associated shim process, which also manages the STDOUT and STDERR streams.


Next, we look at Table \ref{table1} and compare the CPU consumption due to the video streaming and gaming applications. We note that video streaming consumes  much less CPU, in the order of 1/10th,  than the gaming application. Indeed, in order to transmit a stream, a video server simply reads the data and sends it, without any encoding.
\begin{table}[tb]
\centering
\caption{Application load with a fixed number of clients}\label{table1}
  \begin{tabular}[]{ |l|c|c| }
  \hline
  Application & No. of servers & CPU usage [ticks] \\ \hline
  \multirow{4}{*}{FFserver} 
  & 1 server &  392 \\
  & 2 servers & 399 \\
  & 4 servers & 391 \\
  & 8 servers & 365  \\ \hline
  \multirow{4}{*}{Minecraft} 
  & 1 server & 3751 \\
  & 2 servers & 5130 \\
  & 4 servers & 8055 \\
  & 8 servers & 10860 \\ \hline
  \end{tabular}
\end{table}

As a consequence, the  impact of the Docker overhead in the two scenarios is very different. 
This effect is evident  in the plots shown in  Figure \ref{image:fixclient2-3}, which depict the temporal evolution of the Docker overhead and of  the processing load of the applications, in the case with 8 clients and 4 servers. Although the overhead for video streaming is much lower than that for gaming, the overhead of the latter is negligible with respect to the application CPU load, while it accounts for over 10\% of the CPU load in the case of video streaming.

\subsection{Fixed number of servers} \label{fixserver}

\begin{figure}[b!]
\centering
\includegraphics[scale=0.35]{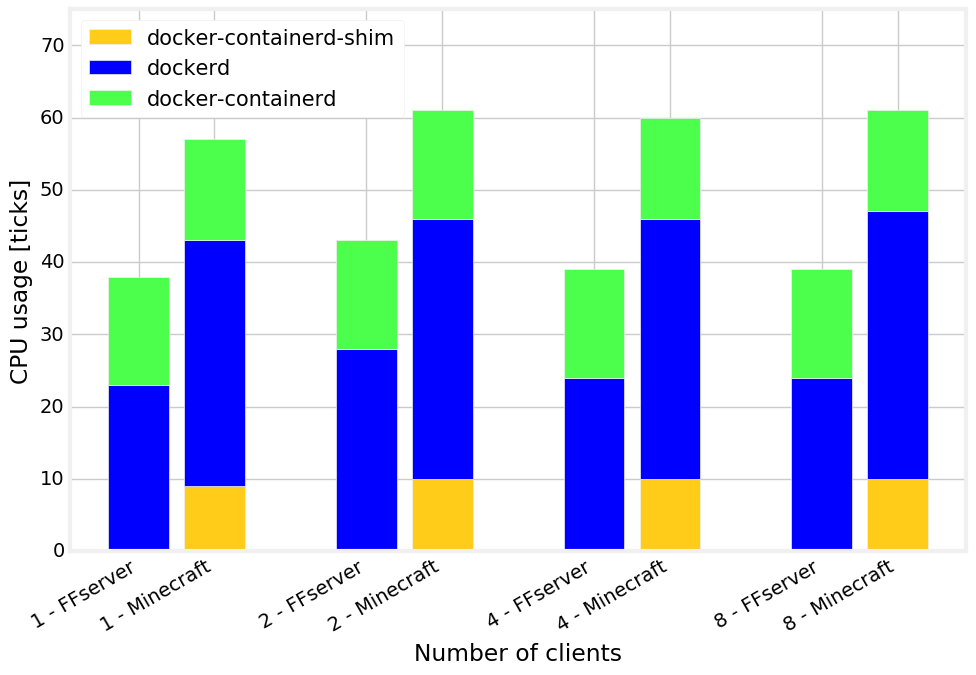}
\caption{Docker overhead CPU consumption when only one server is used and the number of clients varies}\label{img:fixserver1}
\end{figure}

\begin{figure}[b!]
\centering
\includegraphics[scale=0.35]{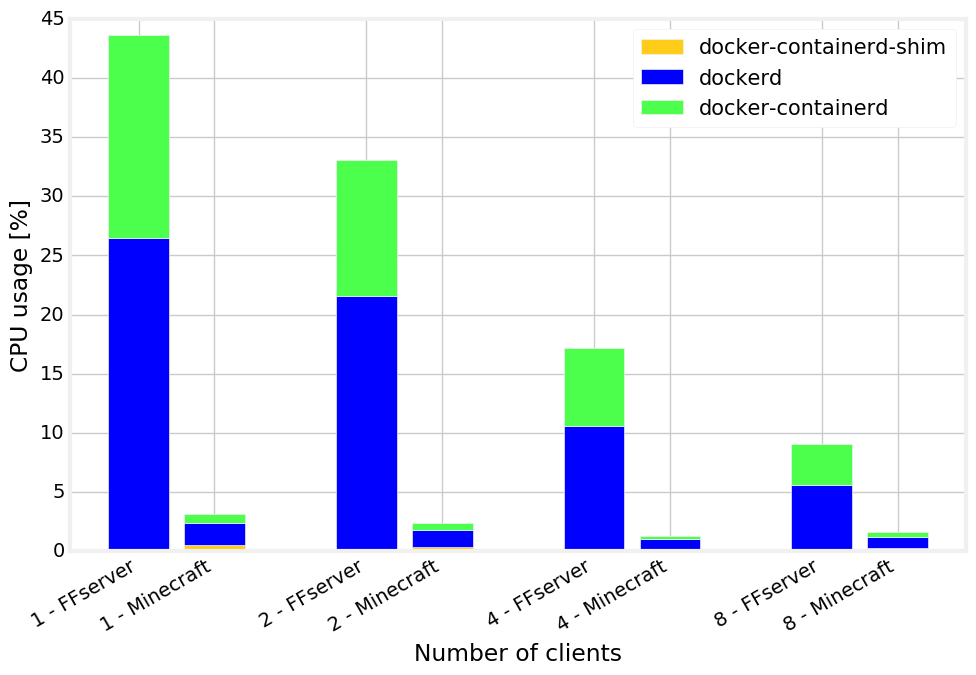}
\caption{CPU consumption (in percentage) due to the Docker overhead, for one server and varying the number of clients}\label{img:fixserver2}
\end{figure}


We now use a single server and vary the number of clients  between 1 and 8, to assess the impact of the number of served flows  on the CPU consumption of Docker processes.

Figure \ref{img:fixserver1} presents the CPU utilization due to the Docker overhead for both our case studies. 

Specifically,  for FFserver we note again that {\tt docker-containerd-shim} (now we have only one server, hence one shim process) does not consume CPU, and that the CPU utilization for both {\tt dockerd} and {\tt docker-containerd} is independent of the number of served flows. 
For Minecraft too results are consistent with what was observed in the previous section, 
however now the CPU usage by  {\tt docker-containerd-shim} and {\tt dockerd} is always constant, i.e., such Docker overhead does not depend on the number of clients served.  This is indeed correct since (i) there are as many shim processes as the number of servers (which in this case is fixed to 1), and (ii)  {\tt dockerd}  manages the running containers, thus, fixed the number of servers, its CPU consumption is constant. 

There is one common point between the two case studies and the different types of tests: the CPU used by {\tt docker-containerd}  is always equal to about 0.05 CPU ticks  per second. We also observed that the execution of this process is unrelated to the normal container execution, indeed, terminating the {\tt docker-containerd} process does not affect the dockerized servers activity.

\begin{table}[t!]
\centering
\caption{Application load with a fixed number of servers}\label{table2}
  \begin{tabular}[]{ |l|c|c| }
  \hline
  Application & No. of clients & CPU usage [ticks] \\ \hline
  \multirow{4}{*}{FFserver} 
  & 1 client &  49 \\
  & 2 clients & 87 \\
  & 4 clients & 188 \\ \hline
  \multirow{4}{*}{Minecraft} 
  & 1 client & 1744 \\
  & 2 clients & 2508 \\
  & 4 clients & 4569 \\ \hline
  \end{tabular}
\end{table}

Table \ref{table2} reports the CPU consumed by the two dockerized servers when 1, 2 and 4 clients are served. 
In the case of FFserver, the overall Docker overhead is about 40 ticks in every test. Thus, 
since the higher the number of clients served, the larger the CPU consumption,  the impact of the Docker overhead sensibly decreases with the increase of the number of clients. In particular,   Figure \ref{img:fixserver2} shows that,  with a single client, the overhead impact  of FFserver is almost 45\%, while it  falls below 10\% for eight clients.
A similar behavior can be observed also for gaming, although less evident as the impact of the Docker overhead  for  Minecraft is already very low in the case of 1 client.

\subsection{Processed data and Docker overhead}\label{data-an} 
\begin{figure}[b!]
\centering
\includegraphics[scale=0.58]{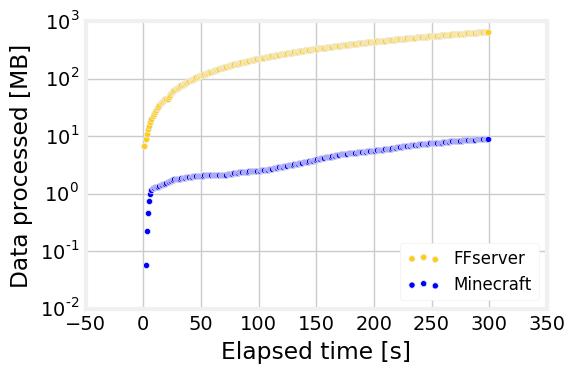}
\caption{Data [MB] processed by the two applications ``1 server - 4 clients'' scenario}\label{img:data}
\end{figure}
We finally establish whether the amount of data processed by the two dockerized servers (i.e., either  transmitted or received) affects the Docker overhead CPU consumption. 

Figure \ref{img:data} represents the amount of data (in logarithmic scale) exchanged between clients and servers in the case of video streaming and gaming applications, for the ``4 clients - 1 server'' scenario. As expected,  the data processed by FFserver is much more than the data transferred by Minecraft, indeed in  our test the video has a size of around 160 MB and  it is transmitted  to four clients. 
Two more facts are worth underlining. First, the quantity of data sent by the video server increases linearly with time, because FFserver does not buffer, being mainly used for live streaming. Second, the Minecraft server processes the largest amount of data early on in the test. Indeed, initially, the server feeds all players information about the virtual world in which they will move. Further data exchanges are due to in-game interactions between server and players, as the latter move around. 
In the ``4 clients - 1 server''  case study, the data transferred after 300~s is about 638~MB for FFserver and 8~MB for Minecraft. We also noted that the Docker CPU consumption remains constant over time and is equal to the value observed before, which was higher for Minecraft than for  FFserver (Figure~\ref{img:fixserver1}).
We can thus conclude that the amount of data processed by the server does not affect the Docker CPU consumption. 

{\bf Takeaway messages}
Our results highlight three important aspects:
\begin{enumerate}
\item throughout our experiments, the Docker overhead ranged between negligible and moderate, thus validating their claim to be a {\em lightweight} containerization solution;
\item such a behavior was observed for two services, namely, video streaming and online gaming, with opposite requirements in terms of CPU load and generated data;
\item in both the case studies, the Docker overhead seems to be independent of the number of clients served, while the number of servers was observed to influence the overhead for Minecraft but not for FFserver.  
\end{enumerate}

\section{Conclusions}
Our analysis focused on Docker overhead in a typical MEC scenario with two different dockerized servers: a video and a multiplayer gaming server. Initially, we looked at different combinations of numbers of servers and clients; then, we analyzed the impact of data processed by the containers.

Different behaviors were observed for each process involved in the running of containerized servers: the CPU consumption of {\tt dockerd}, in the video streaming case, is always constant, while it was found to depend on the number of gaming servers running on the same hardware. The behavior of {\tt docker-containerd} differs: its CPU consumption is independent of the number of clients and servers, in both case studies. As for {\tt docker-containerd-shim}, for each running Minecraft server (regardless of the number of clients served), the CPU consumption is 0.033 ticks per second. Conversely, the \textit{shim} processes of FFserver are always idle.
Finally, we observed that the data processed by each container does not have any impact on the overall Docker overhead CPU consumption.

\section{Acknowledgments}
The authors wish to thank Nigel Brown (\url{http://www.windsock.io}) for his insight about Docker processes. This work has received funding from the 5G-Crosshaul project (H2020-671598).

\bibliographystyle{IEEEtran}
\bibliography{refs.bib}

\begin{thebibliography}{10}
\providecommand{\url}[1]{#1}
\csname url@samestyle\endcsname
\providecommand{\newblock}{\relax}
\providecommand{\bibinfo}[2]{#2}
\providecommand{\BIBentrySTDinterwordspacing}{\spaceskip=0pt\relax}
\providecommand{\BIBentryALTinterwordstretchfactor}{4}
\providecommand{\BIBentryALTinterwordspacing}{\spaceskip=\fontdimen2\font plus
\BIBentryALTinterwordstretchfactor\fontdimen3\font minus
  \fontdimen4\font\relax}
\providecommand{\BIBforeignlanguage}[2]{{%
\expandafter\ifx\csname l@#1\endcsname\relax
\typeout{** WARNING: IEEEtran.bst: No hyphenation pattern has been}%
\typeout{** loaded for the language `#1'. Using the pattern for}%
\typeout{** the default language instead.}%
\else
\language=\csname l@#1\endcsname
\fi
#2}}
\providecommand{\BIBdecl}{\relax}
\BIBdecl

\bibitem{docker:web}
{Docker}. \url{http://docker.com}.

\bibitem{cisco}
{Cisco}.
  \url{http://www.cisco.com/c/en/us/solutions/service-provider/visual-networking-index-vni/index.html}.

\bibitem{game-article}
{The Global Game Market Value}.
  \url{https://newzoo.com/insights/articles/global-games-market-reaches-99-6-}.

\bibitem{dockevaluation}
B.~I. Ismail, E.~M. Goortani, M.~B.~A. Karim, W.~M. Tat, S.~Setapa, J.~Y. Luke,
  and O.~H. Hoe, ``Evaluation of docker as edge computing platform,'' in
  \emph{2015 IEEE Conference on Open Systems (ICOS)}, Aug 2015, pp. 130--135.

\bibitem{AaaS}
T.~Taleb, A.~Ksentini, and R.~Jantti, ``Anything as a service for 5g mobile
  systems,'' \emph{IEEE Network}, vol.~30, no.~6, pp. 84--91, November 2016.

\bibitem{comparison}
W.~Felter, A.~Ferreira, R.~Rajamony, and J.~Rubio, ``An updated performance
  comparison of virtual machines and linux containers,'' in \emph{2015 IEEE
  International Symposium on Performance Analysis of Systems and Software
  (ISPASS)}, March 2015, pp. 171--172.

\bibitem{game}
T.~K\"{a}m\"{a}r\"{a}inen, Y.~Shan, M.~Siekkinen, and
  A.~Yl\"{a}-J\"{a}\"{a}ski, ``Virtual machines vs. containers in cloud gaming
  systems,'' in \emph{Proceedings of the 2015 International Workshop on Network
  and Systems Support for Games}, ser. NetGames '15.\hskip 1em plus 0.5em minus
  0.4em\relax Piscataway, NJ, USA: IEEE Press, 2015, pp. 1:1--1:6.

\bibitem{overlay}
B.~Xie, G.~Sun, and G.~Ma, ``Docker based overlay network performance
  evaluation in large scale streaming system,'' in \emph{2016 IEEE Advanced
  Information Management, Communicates, Electronic and Automation Control
  Conference (IMCEC)}, Oct 2016, pp. 366--369.

\bibitem{ffmpeg}
{FFmpeg}. \url{https://www.ffmpeg.org/}.

\bibitem{vlc}
{VLC}. \url{http://www.videolan.org/vlc/}.

\bibitem{minecraft}
{Minecraft Pocket Edition}. \url{http://www.pocketmine.net/}.

\bibitem{genymotion}
{Genymotion}. \url{https://www.genymotion.com}.

\bibitem{frep}
{FRep}. \url{http://strai.x0.com/frep/}.

\bibitem{cpu}
C.~Ehrhardt, in \emph{CPU time accounting}.\hskip 1em plus 0.5em minus
  0.4em\relax IBM, 2010.

\end{thebibliography}

\end{document}